\documentclass[twocolumn,aps,prc,superscriptaddress,showpacs]{revtex4}
\usepackage{amsmath,bm}%
\usepackage{graphicx}
\usepackage{color}
\begin{document}

\title{Influence of statistical sequential decay  on isoscaling and symmetry energy coefficient in the GEMINI simulation}

\author{ P. Zhou}
  \affiliation{Shanghai Institute of
Applied Physics, Chinese Academy of Sciences,  Shanghai 201800,
China} \affiliation{Graduate School of the Chinese Academy of
Sciences, Beijing 100080,  China}

\author{ W. D. Tian}
\email{tianwendong@sinap.ac.cn}
 \affiliation{Shanghai Institute of
Applied Physics, Chinese Academy of Sciences,  Shanghai 201800,
China}

\author{ Y. G. Ma}
\email{ygma@sinap.ac.cn}  \affiliation{Shanghai Institute of
Applied Physics, Chinese Academy of Sciences,  Shanghai 201800,
China}

\author{ X. Z. Cai}
 \affiliation{Shanghai Institute of
Applied Physics, Chinese Academy of Sciences,  Shanghai 201800,
China}

\author{ D. Q. Fang}
 \affiliation{Shanghai Institute of
Applied Physics, Chinese Academy of Sciences,  Shanghai 201800,
China}

\author{ H. W. Wang}
 \affiliation{Shanghai Institute of
Applied Physics, Chinese Academy of Sciences,  Shanghai 201800,
China}

\date{\today}

\begin{abstract}

Extensive calculations on isoscaling behavior with the
sequential-decay model GEMINI are performed for the mediate-heavy
nuclei in the mass range $A$ = 60-120 at excitation energies  up
to 3 MeV per nucleon. The  comparison between the products after
the first-step decay and the ones after entire-steps decay
demonstrates that there exists strong sequential decay effect on
the final isoscaling parameters and the apparent temperature.
Results show that the apparent symmetry energy coefficient
$\gamma_{app}$ does not reflect the initial symmetry energy
coefficient $C_{sym}$ embedded in the mass calculation in the
present GEMINI model.
\end{abstract}


\maketitle

One of main goals of the isospin physics is to determine the
isospin dependence of the in-medium nuclear effective interactions
and the equation of state (EOS) of isospin asymmetric nuclear
matter or finite nuclei, particularly its isospin-dependent term,
i.e., the density dependence of the nuclear symmetry energy.
Knowledge of nuclear symmetry energy is essential for
understanding not only many problems in nuclear physics, such as
the dynamics of heavy-ion collisions induced by radioactive beams
and the structure of exotic nuclei, but also a number of important
issues in astrophysics, such as the supernova simulation and
neutron star models, which require inputs of  the nuclear equation
of state at extreme values of density and asymmetry
\cite{La91,La00}. Recently impressive progress has been made both
experimentally and theoretically, a number of earlier reviews on
isospin physics with heavy-ion reactions can be found in several
references \cite{Ba05,St05,Li08}.

Symmetry energy could be extracted from heavy-ion collision using
the isoscaling approach \cite{Ts01, Bo02, Sh03, On03, Ve04, Ma04,
Ti05, Ma05, Do06,Ti07}.
 Isoscaling law means that the ratio of isotope
yields $R_{21}(N,Z)=Y_{2}(N,Z)/Y_{1}(N,Z)$, from two similar
reactions, denoted as reaction 1 and 2, which are different only
in their isospin asymmetry,  is found to exhibit an exponential
relationship as a function of the neutron number $N$ and proton
number $Z$ \cite{Ts01}, i. e.

\begin{equation}
R_{21}(N,Z)=\frac{Y_{2}(N,Z)}{Y_{1}(N,Z)}=C\exp(\alpha N+\beta Z),
\label{eq1}
\end{equation}
where $Y_{2}(N,Z)$ and $Y_{1}(N,Z)$ are the fragment yields  from
the neutron-rich and the neutron-deficient reaction, respectively,
$C$ is an overall normalization factor, and $\alpha$ and $\beta$
are fitted parameters. The isoscaling parameter $\alpha$ is
related to the symmetry energy coefficient $C_{sym}$ of EOS in
microcanonical and canonical frames by following relation
\cite{Ts01},

\begin{equation}
\alpha=\frac{4C_{sym}}{T}\left[\left(\frac{Z}{A}\right)_{s1}^{2}-\left(\frac{Z}{A}\right)_{s2}^{2}\right]
\equiv\frac{4C_{sym}}{T}\Delta\left(\frac{Z}{A}\right)_{s}^{2}\label{eq2},
\end{equation}
and
\begin{equation}
\beta=\frac{4C_{sym}}{T}\left[\left(\frac{N}{A}\right)_{s1}^{2}-\left(\frac{N}{A}\right)_{s2}^{2}\right]
\equiv\frac{4C_{sym}}{T}\Delta\left(\frac{N}{A}\right)_{s}^{2}\label{eq3}
\end{equation}
where $Z_{s1}$, $Z_{s2}$, $N_{s1}$, $N_{s2}$, $A_{s1}$, $A_{s2}$
are the charge number, neutron number and mass  numbers of the
sources from the two systems, $T$ is their temperature and
$C_{sym}$ is the symmetry energy coefficient. This relation has
also been evidenced in other model frameworks \cite{Bo02, Sh03,
On03,Ve04, Ma04, Ti05,Ma05, Do06,Ti07}. A great deal of effort has
been devoted to investigate the nuclear symmetry energy and its
density/temperature dependences
\cite{Ba05,Li08,Su08,Nato,Gul,Fang}.

Ideally, primary fragments should be detected right after emission
in order to extract information about the collisions, and Eq.
(\ref{eq2}) is derived based on the primary reaction products
bypassing secondary decays. However, the detected experimental
data are for cold products after the secondary decays from hot
products. Isoscaling has also been reasonably reproduced in the
sequential decay codes \cite{Ts06, Ti07}. However there are still
arguments on the sequential decay effect on isoscaling, some
models show that the effect from sequential decays on isoscaling
is negligible, but some efforts show that sequential decay affects
on the isoscaling parameters, and then distort the extraction of
symmetry energy coefficient $C_{sym}$ \cite{Fe05, Co06}. There are
some issues still keep unsolved or unclear, such as the sequential
decay effect on isoscaling parameters, derived apparent
temperature $T_{r}$ from the experimental measurement and the
isospin evolution of the decaying sources, these factors affect
the extraction of symmetry energy coefficient $C_{sym}$.

The statistical GEMINI model \cite{Ch88} calculates the decay of
compound nuclei by modes of sequential binary decays. The model
employs a Monte Carlo technique to follow the decay chains of
individual compound nuclei through sequential binary decays until
the resulting products are unable to undergo further decay. GEMINI
has been widely used to simulate the hot equilibrium source
de-excitation, or as an $``$afterburner$"$ code to analyze the hot
fragments decay after dynamical simulation \cite{ Ha94,
Ma03,Ch10}. Isoscaling has been investigated by statistical
sequential secondary decay code GEMINI \cite{Ch88}, in which only
the first-step sequential decay was simulated and Eq. (\ref{eq2})
was confirmed for the fragments which are decayed directly from
the initial sources \cite{Ti07}. In the present work, we
investigate the entirely decayed fragments from excited sources,
comparing with the only first-decay fragments from the same
source. The influences of sequential decays  on isoscaling
parameters $\alpha$, $\beta$, and the apparent temperature $T_{r}$
are discussed, and the apparent symmetry energy coefficient
$\gamma_{app}$ is extracted.

The detailed description of GEMINI code can be found in
\cite{Ch88, Ti07}, the same configuration and parameters of the
GEMINI code were adopted as in Ref. \cite{Ti07}.  Several pairs of
equilibrated sources are considered at various initial excitation
energy $E_{ex}$=1.0, 1.4, 2.0, 2.4, and 3.0 MeV/nucleon.  We
selected source pairs with the same proton number $Z_{s}$ but
different mass number $A_{s}$ to systematically study the
isoscaling behavior. In this case,  possible effects of different
magnitudes of Coulomb interaction on isotopic distributions are
avoided. The equilibrated source pairs are chosen in different
mass region and system isospin asymmetry $N/Z$. Two groups of the
source pairs have been used: (1)$Z_{s}$=50 with $A_{s}$=100, 105,
110, and 115, respectively; (2)$Z_{s}$=30 with $A_{s}$=60, 63, 66,
and 69, respectively. Following literature the index $``$2$"$
denotes more neutron-rich system as widely used in convention, and
index $``$1$"$ denotes the the more neutron-deficient system. In
our previous work \cite{Ti07}, the statistical decay stops after
one particle emitted from the source, that was called the
$``$first-step$"$ decay in this paper, which is a simple picture
and the decay procedure can be expressed definitely and clearly,
isoscaling has been confirmed for the first-step decay products in
detail, and the reasonability of extracting symmetry energy
coefficient $C_{sym}$ from the simulation results via experimental
analysis technique. But the first-step decay was not the real
case, experiments measure the final products after multi-step
decays until no fragments produced or gamma rays emitted, which
was called $``$entire-steps$"$ decay in this paper.

\begin{figure}
\vspace{1.0truein}
\includegraphics[scale=0.25]{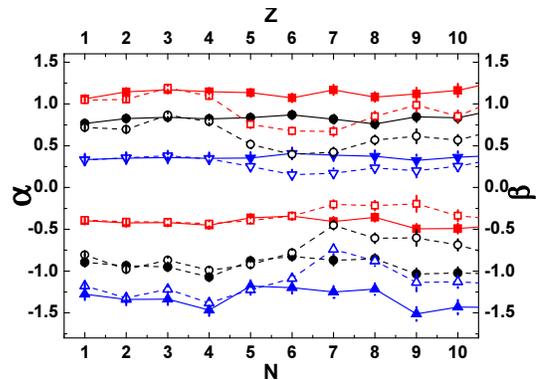}
\vspace{-0.9truein} \caption{\footnotesize (Color online)
Comparisons of isoscaling parameters $\alpha$ (positive values) or
$\beta$ (negative values) as a function of  $Z$ or $N$ from source
pairs of $Z_{s}=50$ at excitation energies $E_{ex}$ = 2.4
MeV/nucleon. All solid symbols represent the results for only the
first-step secondary decay products, open symbols for entire-step
secondary decay products. Symbols in the figure correspond to
$Y_{A_{s}=115}$/$Y_{A_{s}=100}$ (up-triangles),
$Y_{A_{s}=110}$/$Y_{A_{s}=100}$ (circles),
$Y_{A_{s}=110}$/$Y_{A_{s}=105}$ (down-triangles), respectively.}
\label{isoscaling}
\end{figure}

Isoscaling is analyzed from the emitted light fragments in above
both cases, namely first-step decay only and entire-steps decay
chains for all the simulated systems. As an example shown in Fig.
\ref{isoscaling} the comparison of the isoscaling parameters
$\alpha$ and $\beta$ is plotted as a function of emitted light
fragment proton number $Z$ and neutron number $N$, between the
first-step decay and entire-steps decay fragments for the source
pairs of $Z_{s}$ = 50. As we can see, isoscaling parameters
$\alpha$ and $\beta$ are essentially independent on  $Z$ or $N$ of
fragments, for both first-step and entire-steps decay fragments,
 especially for $Z \leq 4$ or $N \leq 5$. It has been
evidenced that in the first-step decay case, the probability of
producing a cluster with a given $Z$ and  $A$ at $T$ depends
exponentially on the free energy of that cluster, $ F(Z,A,T)$, in
GEMINI simulation, the cluster free energies depend on the
strength of the symmetry term in the liquid-drop energy through
the Eq. (\ref{eq2}) with $C_{sym} \approx$ 24 MeV\cite{Ti07}. If
the entire-steps sequential secondary decay is included, the
source isospin $N/Z$ and temperature $T$ as well as isoscaling
parameters vary after each step of decay. Similarly, in previous
study, time dependence of isoscaling parameters have been
discussed in molecular dynamics model \cite{Do06}, indicating that
the final values of these parameters could be related to the last
part of the reaction where the fragments finish cooling by
particle evaporation. In present GEMINI model, fragment yields are
strongly expected degraded after entire-steps decay, so do the
isotopic yield ratios. It is not surprising that in Fig.
\ref{isoscaling} that parameters $\alpha$ and $\beta$ extracted
from isotopic yield ratios of the final emitted light fragments
show discrepancy, especially for the intermediate mass fragments
like $Z\geq5$ in some cases, since those heavier fragments
experience strong multi-step decay and feeding-down effects.
Finally, $\alpha$ and $\beta$ change a lot comparing with the only
first-step decay case, and the fluctuation increases in the
entire-steps decay case. Average $\alpha$ and $\beta$ values over
fragments $Z$ and $N$ are used in following discussions to observe
the overall property.

\begin{figure}
\hspace{-0.45truein}
\includegraphics[scale=0.66]{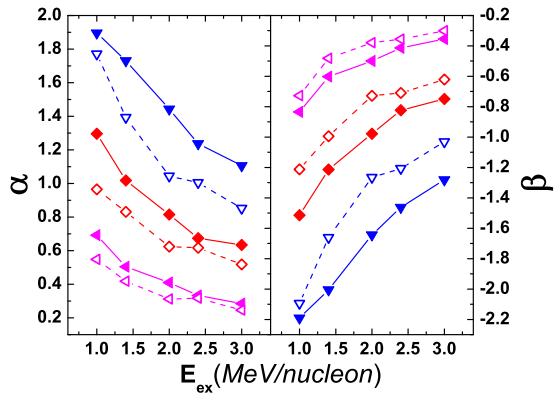}
\vspace{-0.3truein} \caption{\footnotesize (Color online)
Comparisons  of isoscaling parameters $\alpha$ (left panel) and
$\beta$ (right panel) as a function of the source excitation
energy from source pairs of $Z_{s}$=30. All solid symbols
represent only the first-step secondary decay products, open
symbols are entire steps secondary decay products. Symbols in the
figure correspond to 
$Y_{A_{s}=69}$/$Y_{A_{s}=60}$ (down-triangles),
$Y_{A_{s}=69}$/$Y_{A_{s}=63}$ (diamonds),
$Y_{A_{s}=69}$/$Y_{A_{s}=66}$ (left-triangles).}
\label{alpha-beta}
\end{figure}

In Fig. \ref{alpha-beta} we show the comparison of isoscaling
parameters $\alpha$ and $\beta$ as a function of  excitation
energy of sources from different source pairs. If the entire-steps
decay chains are included, which are depicted  by the open symbols
in Fig. \ref{alpha-beta}, $\alpha$ and $|\beta|$ values  show
significant decrease. This reduction is about 20$\%$ in average,
consistent with the result in Ref. \cite{Ts06, Fe05} for the
$C_{sym} \approx$ 25MeV case, but the excitation energy dependent
trend of $\alpha$ and $\beta$  does not change.

As already discussed, the isoscaling parameter $\alpha$ is related
to  the symmetry energy coefficient $C_{sym}$ of the nuclear
binding energy through Eq. (2) and (3), this relation provides a
direct link between the measurable quantities and the nuclear
symmetry energy coefficient. It should be noticed that the
parameters $\alpha$ and $\beta$ refer to the hot primary
fragments, which have to  undergo sequential decays into the cold
fragments. It was assumed that the secondary decay on the yield of
a specific isotope is similar for the two reactions, thus the
effect of the sequential decays on $R_{21}(N, Z)$ is small and
that $R_{21}(N, Z)$ reflects the properties of the primary source.
In our present study, the first-step statistical sequential decay
process stems from a fixed initial source with definite excitation
energy (temperature) and isospin asymmetry, and it has been
verified by Eq. (\ref{eq2}) and  Eq. (\ref{eq3}), to reflect the
link between the measurable quantity $R_{21}(N, Z)$ and the
symmetry energy coefficient $C_{sym}$.

It has been found in Fig. \ref{isoscaling}, the isoscaling
behavior still presents after considering the entire-steps decay
chains. But in this case,  the isoscaling parameters $\alpha$ and
$\beta$ decrease comparing with the only first-step decay case  as
seen in Fig. \ref{alpha-beta}. In the statistical sequential
decay, the source temperature $T$ and isospin asymmetry $N/Z$ also
change after each step of the decays, thus the parameters $T$ and
$\Delta(Z/A)_{s}^{2}$ (or $\Delta(N/A)_{s}^{2}$) varies during the
sequential decay process, where many intermediate sources are
different from the initial source.

\begin{figure}
\includegraphics[scale=0.3]{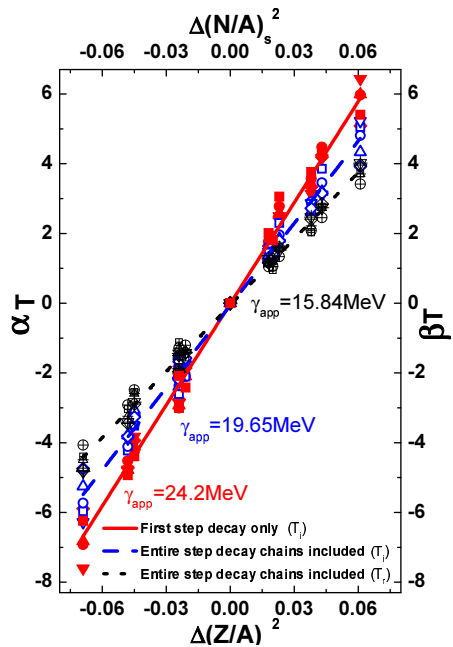}
\vspace{-0.0truein} \caption{\footnotesize (Color online) $\alpha
T$ ($\beta T$) as a function of the source isospin difference
$\Delta(Z/A)_{s}^{2}$ ($\Delta(N/A)_{s}^{2}$) from source pairs of
$Z_{s}$ = 30. All solid symbols represent only the first-step
secondary decay case, open symbols without/with cross represent
the entire-steps secondary decay products with $T_i$ and $T_r$,
respectively. Symbols in the figure correspond to excitation
energies $E_{ex}/A$ = 1.0 (squares), 1.4 (circles), 2.0
(up-triangles), 2.4 (down-triangles) and 3.0 MeV (diamonds),
respectively. The solid, dash and dot lines are the linear fitting
for the above three cases which gives the apparent symmetry energy
coefficient $\gamma_{app}$ = 24.2, 19.65 and 15.84 MeV,
respectively.} \label{Fig-at1}
\end{figure}

To explore the validity of Eq. (\ref{eq2}) and (\ref{eq3}) in the
entire-steps statistical  sequential decay, we plot $\alpha T$ as
a function of $\Delta(Z/A)_{s}^{2}$ and $\beta T$ as a function of
$\Delta(N/A)_{s}^{2}$ in Fig.~\ref{Fig-at1}.
 For the first-step only decay, data points depicted
by the solid symbols, with using the initial source temperature
$T_i$ which was calculated by input excitation energy as shown in
Table I  and isoscaling parameters after the first step decay. The
linear fit (the solid line)  gives a symmetry energy coefficient
$C_{sym}$ = 24.2 $\pm$ 0.3MeV. To investigate the decay effect on
symmetry energy coefficient, data points from the entire-steps
decays are also plotted in Fig. \ref{Fig-at1} as shown by the open
symbols. As we have mentioned for the entire-steps decay chains,
there are many intermediate-state sources with different
temperature $T$ and isospin asymmetry $Z/A$ or $N/A$.
Nevertheless, the final isoscaling parameters $\alpha$ and $\beta$
still show similar rules as the first-step decay only case, i.e.
$\alpha T$ ($\beta T$) and $\Delta(Z/A)_{s}^{2}$
($\Delta(Z/A)_{s}^{2}$) can be still fitted by another linear
function, namely
\begin{equation}
\alpha=\frac{4\gamma_{app}}{T}\left[\left(\frac{Z}{A}\right)_{s1}^{2}-\left(\frac{Z}{A}\right)_{s2}^{2}\right]
\equiv\frac{4\gamma_{app}}{T}\Delta\left(\frac{Z}{A}\right)_{s}^{2}\label{eq4},
\end{equation}
and
\begin{equation}
\beta=\frac{4\gamma_{app}}{T}\left[\left(\frac{N}{A}\right)_{s1}^{2}-\left(\frac{N}{A}\right)_{s2}^{2}\right]
\equiv\frac{4\gamma_{app}}{T}\Delta\left(\frac{N}{A}\right)_{s}^{2}\label{eq5}.
\end{equation}
which gives  the apparent symmetry energy coefficient
$\gamma_{app}$ = 19.65 $\pm$ 0.25 MeV if the same T$_i$ are used.

\begin{table}
\caption{\label{tab1} Initial temperature ($T_i$) and final-state
apparent temperature ($T_r$) in different systems ($Z_s$=30, $A_s$
= 60 or 63, 66 or 69.}
\begin{ruledtabular}
\begin{tabular}{cccccc}
\multicolumn{2}{c} {$E_{ex}$}& $T_i$ & $T_{r}$ &$T_{i}$
&$T_{r}$ (MeV)  \\
&$A_s$&  60/63 & 60/63 & 66/69 & 66/69
\\
\hline
& 1.0 & 2.8/2.9  & 2.3/2.4 & 2.9/2.9 & 2.4/2.4\\
 & 1.4 & 3.4/3.5 & 2.8/2.9& 3.5/3.5 & 2.9/3.0 \\
   & 2.0 & 4.1/4.2  & 3.5/3.6& 4.2/4.2 & 3.6/3.6\\
  & 2.4 & 4.5/4.6 & 3.8/3.9& 4.6/4.7& 3.9/4.0 \\
 & 3.0 & 5.1/5.1 & 4.3/4.4& 5.2/5.3& 4.4/4.5\\
\end{tabular}
\end{ruledtabular}
\end{table}

In fact, for the source decays with entire steps, the
intermediate-state sources have different isospin asymmetry $N/Z$
ranging from initial isospin asymmetry to the stable line or the
evaporation attract line \cite{Ch98} and different temperature $T$
ranging from initial temperature to zero. In this case,
principally both $T$ and $\Delta(Z/A)_{s}^{2}$
($\Delta(Z/A)_{s}^{2}$) need to be corrected to reflect the
intermediate sources. In the simulation, the initial source
temperature $T$ can be calculated, and the intermediate source
tracing the sequential decay chains can also be performed. From an
experimental point of view, temperature and isospin asymmetry of
the intermediate source can be extracted from evaporation products
which reflects the whole decay chains.

Traditionally, temperature can be extracted from the measurements
of spectral slopes or double isotopic ratios at lower energies
\cite{Al85,Ma97}. In the present work, initial temperature $T_{i}$
are calculated directly in the GEMINI code by the input excitation
energy \cite{Ti07}, and the final-state temperature $T_{r}$ can be
obtained by
 the neutron and proton spectra fitting when the
entire-steps decay chains are included in the GEMINI calculation.
The results are displayed in TABLE. I.

When the temperature $T_{r}$ is used in Eq. (\ref{eq4}) and
(\ref{eq5}) to fit the linear slope parameter $\gamma_{app}$, it
leads to the reduction of the parameter $\gamma_{app}$ as shown in
Fig. \ref{Fig-at1} (short dashed line). Its slope gives an
apparent symmetry energy coefficient $\gamma_{app}$ = 15.84 $\pm$
0.18 MeV, which is one-third reduction comparing with the only
first-step decay case. In this context, we should be careful to
use the apparent symmetry energy derived directly from the final
fragments which could be distorted due to the multi-step
sequential decays. Of course, the present results are specific for
the use of GEMINI to describe the secondary decay, i.e. they  may
depend on the details of the sequential decay code.

In summary, we performed the isoscaling analysis for both light
fragments from only the first-step decay and the entire-steps
decay chains with GEMINI code, it is found that isoscaling can
still be observed and the Eq. (4) and (5) which are used to
extract the symmetry energy coefficient also work after the
entire-steps decay is taken into account. However, the statistical
sequential decay leads to the decreasing of isoscaling parameters
$\alpha$ and $\beta$ as well as temperature. Therefore, the
reduced ({apparent}) source temperature together with the reduced
isoscaling parameters leads to a smaller symmetry energy parameter
$\gamma_{app}$  in comparison with the initial symmetry energy
coefficient $C_{sym}$ which is constrained from the only
first-step statistical decay calculation.  From the present GEMINI
model calculations, we shall carefully consider the multi-step
sequential decay effect on the extraction of  the symmetry energy
coefficient via the final cold products.

This work is supported in part by NSFC under contract No.s
11035009, 10875167, 1097907 and 11005140, the 973-Program under
contract No. 2007CB815004, and the Shanghai Development Foundation
for Science and Technology under contract No. 09JC1416800.



\begin{thebibliography}{50}

\bibitem{La91} J. M. Lattimer {\it et al.}, Phys. Rev. Lett. {\bf 66}, 2701 (1991).
\bibitem{La00} J. M. Lattimer and M. Prakash, Phys. Rep. {\bf 333}, 121 (2000).
\bibitem{Ba05} V. Baran {\it et al.}, Phys. Rep. {\bf 410}, 335 (2005).
\bibitem{St05} A. W. Steiner, M. Prakash, J. M. Lattimer, P.J. Ellis, Phys. Rep. {\bf 411}, 325 (2005).
\bibitem{Li08} B. A. Li, L. W.  Chen, C. M. Ko, Phys. Rep. {\bf 464}, 113 (2008).
\bibitem{Ts01} M. B. Tsang {\it et al.}, Phys. Rev. Lett. {\bf 86}, 5023 (2000).
\bibitem{Bo02} A. S. Botvina, O. V. Lozhkin, and W. Trautmann, Phys. Rev. C {\bf 65}, 044610 (2002).
\bibitem{Sh03} D. V. Shetty et al., Phys. Rev. C 68, 021602(R) (2003).
\bibitem{On03} A. Ono {\it et al.}, Phys. Rev. C {\bf 68}, 051601 (2003).
\bibitem{Ve04} M. Veselsky, G. A. Souliotis, and S. J. Yennello, Phys. Rev. C {\bf 69}, 031602(R)
(2004).
\bibitem{Ma04} Y. G. Ma {\it et al.}, Phys. Rev. C {\bf 69}, 064610 (2004).
\bibitem{Ti05} W. D. Tian {\it et al.}, Chin. Phys. Lett.  {\bf 22}, 306 (2005).
\bibitem{Ma05} Y. G. Ma {\it et al.}, Phys. Rev. C {\bf 72}, 064603 (2005).
\bibitem{Do06} C. O. Dorso {\it et al.}, Phys. Rev. C {\bf 73}, 044601 (2006).
\bibitem{Ti07} W. D. Tian {\it et al.}, Phys. Rev. C {\bf 76}, 024607 (2007).
\bibitem{Su08} Q. M. Su {\it et al.}, Chin. Phys. Lett. {\bf 25}, 200 (2008).
\bibitem{Nato}J. B. Natowitz {\it et al.}, Phys. Rev. Lett. {\bf 104}, 202501 (2010).
\bibitem{Gul} G. Lehaut, F. Gulminelli, and O. Lopez, Phys. Rev. Lett. {\bf 102}, 142503 (2009).
\bibitem{Fang}D. Q. Fang  {\it et al.}, J. Phys. G {\bf 34}, 2173 (2007).
\bibitem{Ts06} M. B. Tsang {\it et al.}, Eur. Phys. J. A {\bf 30}, 129 (2006).
\bibitem{Fe05} A. Le F\`{e}vre {\it et al.}, Phys. Rev. Lett. {\bf 94}, 162701 (2005).
\bibitem{Co06} M. Colonna and M. B. Tsang, Eur. Phys. J. A {\bf 30}, 165 (2006).
\bibitem{Ch88} R. J. Charity {\it et al.}, Nucl. Phys. A {\bf 483}, 371 (1988);\\
               R. J. Charity, computer code GEMINI, see http://www. chemistry.wustl.edu/$^{\sim}$rc/.
\bibitem{Ha94} K. Hagel {\it et al.}, Phys. Rev. C {\bf 50}, 2017 (1994).
\bibitem{Ma03} Y. G. Ma {\it et al.}, Phys. Rev. C {\bf 65}, 051602(R) (2002).
\bibitem{Ch10} R. J. Charity£¬Phys. Rev. C {\bf 82}, 014610 (2010).
\bibitem{Ch98} R. J. Charity, Phys. Rev. C {\bf 58}, 1073 (1998).
\bibitem{Al85} S. Albergo{\it  et al.}, Il Nuovo Cimento A {\bf 89}, 1 (1985).
\bibitem{Ma97}Y. G. Ma {\it et al.}, Phys. Lett. B {\bf 391}, 41 (1997).

\end{thebibliography}
\end{document}